\documentstyle[12pt]{article}
 \newcommand{\bea}{\begin{equation}}
 \newcommand{\eea}{\end{equation}}
 \newcommand{\ber}{\begin{eqnarray}}
 \newcommand{\eer}{\end{eqnarray}}
\begin{document}
 \title{\bf Coupled non-equilibrium growth equations:
 Self consistent mode coupling using vertex
 renormalization}
 \author{${}^{1}$ Amit Kr. Chattopadhyay,${}^{2}$
 Abhik Basu,\\
 ${}^{1}$ Jayanta K. Bhattacharjee,\\\\
 ${}^{1}$ Dept. of Theoretical Physics,\\
 Indian Association for the Cultivation of Science,\\
 Jadavpur, Calcutta 700 032\\\\
 ${}^{2}$ Dept. of Physics,\\
 Indian Institute of Science, Bangalore 560 012}
 \maketitle
 \date{}
 \begin{abstract}
 We find that studying the simplest of the coupled
 non-equilibrium growth 
 equations of Barabasi by self-consistent mode
 coupling requires the use
 of dressed vertices. Using the vertex
 renormalization, we find a roughening exponent which
 already in the leading order is quite close to the
 numerical value.
 \end{abstract}
 PACS number(s):05.10.Gg,05.40.-a,05.70.-a,64.60.Ht
 \newpage
 Models of interfacial growth have attracted a
 tremendous amount of attention 
 since the pioneering work of Kardar, Parisi, Zhang
 (KPZ) [1,2]. A variety of interesting issues are
 associated with the KPZ equation and they have given
 rise to a variety of novel techniques [3]. Among the
 first analytic techniques used to tackle the KPZ
 system were the dynamic renormalization group (DRG)
 [4] and the
 self-consistent mode coupling scheme (SCMC) [5,6].
 An important variant of the 
 KPZ system was introduced by Ertaas and Kardar [7]
 and Barabasi [8]. This variant consisted of two
 coupled fields (as opposed to one field in KPZ) and
 is useful for studying the effects of a second
 non-equilibrium field on the growing interface. In
 these coupled field problems DRG has been employed,
 as also numerical techniques. One does not always
 get a stable fixed point with the DRG analysis 
 which may sometimes indicate a failure of the
 perturbation scheme or may
 indicate a basic instability of the system. It is
 interesting to note that 
 in many cases the exponents coming from the one-loop
 DRG analysis are not 
 in very good agreement with the numerical analysis.
 This is exemplified in the 
 simplest situation treated by Barabasi - an
 essentially linear system coupled according to
 
 \bea
 \frac{\partial \phi}{\partial t} =
 \Gamma_1\:\frac{\partial^2 \phi}{\partial
 x^2}\:+\:N_1
 \eea
 
 \bea
 \frac{\partial \psi}{\partial t} =
 \Gamma_2\:\frac{\partial^2 \psi}{\partial
 x^2}\:+\:\lambda\:\frac{\partial \phi}{\partial x}
 \frac{\partial \psi}{\partial x}\:+\:N_2
 \eea
 
 \hspace{2cm}
 with $ <N_{1,2}(x_1,t_1)\:N_{1,2}(x_2,t_2)>\:=\:2
 D_{1,2}\:\delta(x_1-x_2)\:
 \delta(t_1-t_2) $.
 \vspace{1cm}
 \par
 The field $ \phi $ satisfies the Edwards-Wilkinson
 equation and the field $ \psi $ is coupled linearly
 via a gradient
 coupling to the $ \phi $-field. While the
 Edwards-Wilkinson model can be exactly solved, this
 is not true for eqn.(2)
 because of the multiplicative noise (note that $
 \phi $ is a random field). The DRG recursion
 relations in this case yield
 for the roughening exponent $ \alpha $ of the $ \psi
 $-field, the value $ \alpha=5/6 $ while the
 numerical value of $ \alpha $
 is nearly 0.68. The dynamical exponent $ z $ of the
 $ \psi $-field is found to be 2. 
 Thus, in this case the dynamic exponent for both $ \phi $ and $ \psi $ fields is found to be 2. We will call this 
"extended" dynamic scaling i.e. the time scale is independent of the nature of the field [9]. As it turns out, this is the only situation for this case. However this need not always be so. In another model considered by Ertaas and Kardar and Barabasi,

 \bea
 \frac{\partial \phi}{\partial t} =
 \Gamma_1\:\frac{\partial^2 \phi}{\partial
 x^2}\:+\:\lambda_1\:{(\frac{\partial \phi}{\partial x})^2}\:+\:N_1
 \eea
 
 \bea
 \frac{\partial \psi}{\partial t} =
 \Gamma_2\:\frac{\partial^2 \psi}{\partial
 x^2}\:+\:\lambda\:\frac{\partial \phi}{\partial x}
 \frac{\partial \psi}{\partial x}\:+\:N_2
 \eea
 
 there are two possibilities: \\
i)$ z_{\phi}=z_{\psi}=3/2 $, this is the extended dynamic scaling and is found to be the correct situation for $ \lambda > 0 $ with $ \lambda_1 > 0 $,\\
ii) $ z_{\phi}=3/2 $, but $ z_{\psi} =2 $, this situation is obtained for $ \lambda < 0 $ with $ \lambda_1 > 0 $ and can be described as "weak" scaling [9]. For problems involving two or more coupled fields, one needs to differentiate between "extended" and "weak" scaling.
\par
 In the one-dimensional KPZ, the perturbative DRG is
 exact (due to the existence of a
 fluctuation-dissipation relation),
 but this is not true for the coupled system in one
 dimension.The self-consistent mode coupling (SCMC)
 which has been 
 reasonably succesful for the KPZ, has never been
 attempted in the coupled system. In this note, we
 apply the SCMC
 to the coupled system to see if it is a
 quantitatively better scheme than the perturbative
 DRG. In the process, we find
 something quite unusual.
 In all known situations, SCMC has been succesful in
 cases where the vertex is not
 renormalized. This, in contrast, is a situation
 where the momentum dependence of the dressed vertex
 is absolutely 
 essential. This is what makes the application of
 SCMC interesting in this problem and should act as a
 prototype for
 situations where dressed vertices are unavoidable.
 Writing eqns. (1) and (2) in momentum space, we have

 \bea
 \dot{\phi(k)} = -\Gamma_1\:k^2\:\phi(k)\:+\:N_1(k)
 \eea
 
 \bea
 \dot{\psi(k)} =
 -\Gamma_2\:k^2\:\psi(k)\:-\:\lambda\:\sum_{p}\:p(k-p)\:\phi(p)
 \psi(k-p)\:+\:N_2(k)
 \eea
 
 with $ <N_{1,2}(k_1,\omega_1)
 N_{1,2}(k_2,\omega_2)>\:=\:2
 D_{1,2}\:\delta(k_1+k_2)\:\delta(\omega_1+\omega_2)
 $.i
 \vspace{1cm}
 \par
 The basic elements of the calculation are the
 Green's functions $ G_{\phi}(k,\omega) $ and $
 G_{\psi}(k,\omega) $, the
 correlation functions $ C_{\phi}(k,\omega) $ and $
 C_{\psi}(k,\omega) $ and the vertex function $
 \Lambda(k,q,k-q) $.
 Obviously $ G_{\phi} $ and $ C_{\phi} $ are exactly
 known and are given by 
 
 \bea
 G^{-1}_{\phi}(k,\omega) = -i \omega\:+\:\Gamma_1 k^2
 \eea
 
 \bea
 C_{\phi}(k,\omega) = \frac{2
 D_1}{\omega^2+{\Gamma_1}^2 k^4}
 \eea
  
 while for the $ \psi $-field
 
 \bea
 G^{-1}_{\psi}(k,\omega) = -i \omega\:+\:\Gamma_2
 k^2\:+\:\Sigma(k,\omega)
 \eea
 
 \bea
 C_{\psi}(k,\omega) = \frac{2
 D_2}{\omega^2+{\Gamma_2}^2 k^4}\:+\:{\mid
 G_{\psi}(k,\omega) \mid}^2 F(k,\omega)
 \eea
 
 and
 
 \bea
 \Lambda(k,p,k-p) = \lambda\:+\:\Lambda(k,p)
 \eea
 
 \par
 The self-energy $ \Sigma(k,\omega) $ is found at the dressed one level to be given by

\begin{eqnarray}
\Sigma(k,\omega) &=& {\lambda}^2\:\int\:\frac{dp}{2\pi} \frac{d{\omega}^{\prime}}{2\pi}\:k{p^2}(k-p)\:C_{\phi}(p,{
\omega}^{\prime}) G_{\psi}(k-p,\omega-{\omega}^{\prime}) \nonumber\\
&=&\frac{{\lambda}^2 D_1}{\Gamma}\:\int\:\frac{dp}{2\pi}\:
\frac{k(k-p)}{-i\omega+\Gamma_1 p^2+\Sigma(k-p)+\Gamma_2 (k-p)^2}
\end{eqnarray}

where we have used eq.(8) and eq.(9) in the Lorentzian approximation, i.e. during the frequency convolution, $ \Sigma_{\psi}(k,\omega) $ has been replaced by its zero frequency form.
\par
Our first observation is that within the extended dynamic scaling, we expect
$ z_{\psi}=2 $. We need to examine if this is self-consistent. Setting
$ \Sigma(k)=\Gamma k^2 $, we have

\bea
\Gamma k^2 = \frac{{\lambda}^2 D_1}{\Gamma_1}\:\int\:\frac{dp}{2\pi}\:\frac{k(k-p)}{\Gamma_1 p^2+ \tilde{\Gamma_2} (k-p)^2}
\eea

\hspace{2cm}
where $ \tilde{\Gamma_2}=\Gamma_2+\Gamma $.\\
The long wave length
property ($ k \rightarrow 0 $) of the integral on the right hand side is best
seen by
changing to the symmetric variables $ p^{\prime}=\frac{-k}{2}+p $ which gives 
the o($ k^2 $) contribution of the integral to be $ k^2\:\frac{{\lambda}^2 D_1}{\Gamma_1}\:\int\:\frac{dp^{\prime}}{2\pi}\:\frac{3\Gamma_1-\tilde{\Gamma_2}}{(\Gamma_1+\Gamma_2)}\:\frac{1}{{p^\prime}^2} $. This integral is divergent and needs to be cut-off at o(k), which spoils the $ k^2 $ behaviour. The only way this can be prevented is by setting $ 3 \Gamma_1=\tilde{\Gamma_2} $, which makes the o($ k^2 $) contribution of $ \Sigma $ vanish, i.e. implies $ \Gamma=0 $ and this establishes

\bea
3 \Gamma_1 = \Gamma_2
\eea

which is in exact agreement with the earlier work of Barabasi.
\par
 We now discuss the correlation function. The diagram
 with bare vertex is shown in Fig.1a and leads to 
 
 \ber
 C_{\psi}(k,\omega) &=& \frac{2
 D_2}{\omega^2+{\Gamma_2}^2 k^4}\:+\:{\mid
 G_{\psi}(k,\omega) \mid}^2 \lambda^2\:\int\:
 \frac{dp}{2\pi} \frac{d\omega^{\prime}}{2\pi}
 \nonumber\\
 & & p^2 (k-p)^2 \: C_{\phi}(p,\omega^{\prime})
 C_{\psi}(k-p,\omega-\omega^{\prime}) 
 \end{eqnarray}
 
 We now assume the scaling form   
 
 \bea
 C_{\psi}(k,\omega) =
 \frac{D_{\psi}}{k^{3+2\alpha}}\:f(\omega/k^2)
 \eea
 
 which is consistent with the equal time correlation
 function, $
 \int\:\frac{d\omega}{2\pi}\:C_{\psi}(k,\omega) $
 being
 $\ k^{-1-2\alpha} $. In the absence of $ \lambda $,
 $ \alpha =1/2 $ and the extra roughness produced by
 this added noise
 is expected to raise $ \alpha $ beyond $ 1/2 $. Our
 expectation, then is that the second term will
 dominate in eqn.(15).
 The power count of the second term in eqn.(15) shows
 that $ C_{\psi}(k,\omega) \sim k^{-4-2\alpha} $
 which cannot match
 the power count of the left hand side for any value
 of $ \alpha $ and hence a self-consistent
 formulation requires the 
 vertex to acquire a momentum dependence. Dressing
 the vertex leads to the diagram in Fig.1b. Dropping
 the first term 
 on the right hand side of eqn.(15) and dressing the
 vertex in the second leads to 
 
 \ber
 C_{\psi}(k,\omega) &=& {\mid G_{\psi}(k,\omega)
 \mid}^2\:\lambda\:\int\:\frac{dp}{2\pi}\frac{d\omega^{\prime}}{2\pi}
 p^2(k-p)^2\:\Lambda(k,p,k-p) \nonumber\\  
 & & C_{\phi}(p,\omega^{\prime})
 C_{\psi}(k-p,\omega-\omega^{\prime})
 \end{eqnarray}
 
 Since, we are interested in the $ k \rightarrow 0 $
 property of $ C_{\psi}(k,\omega) $, the vertex that
 we need is
 $ Lim_{k \rightarrow 0} \Lambda(k,p,k-p) $ and if in
 this limit the vertex has the form $ \Lambda_0 p $
 where $ \Lambda_0 $ is a constant, then the
 self-consistency in power counting is restored. The
 consistency of the amplitude is assured
 if (we evaluate the integral in eqn.(17) in the
 leading approximation [12] of $ k \rightarrow 0 $)
 
 \ber
 1 &=& \frac{\lambda D_1}{9
 {\Gamma_1}^2}\:\int\:\frac{dp}{2\pi}\:\frac{(k-p)^2}{{\mid
 k-p \mid}^{1+2\alpha}}\:
 \frac{\Lambda(k,p,k-p)}{(k-p)^2+\frac{1}{3}p^2}
 \nonumber\\
 & \simeq & \frac{\lambda \Lambda_0 D_1}{9
 {\Gamma_1}^2}\:\int\:\frac{dp}{2\pi}\:\frac{1}{\frac{4}{3}
 p^{2\alpha}}
 \nonumber\\
 &=& \frac{\lambda \Lambda_0 D_1}{12 {\Gamma_1}^2}
 \frac{1}{2 \alpha - 1}
 \end{eqnarray}
 
 \par
We note in passing that the above momentum dependence of the vertex does not alter the conditions of eqn.(14).
 The self consistent equation for the vertex is shown
 in Fig.2. Clearly
 
 \ber
 \Lambda(k,q,k-q) &=& \lambda\:\int\:\frac{dp}{2\pi}
 \frac{d\omega}{2\pi}\:p(p-q)(k-p)^2\: 
 G_{\psi}(p,\omega) G_{\psi}(p-q,\omega) \nonumber\\
 & & C_{\phi}(k-p,\omega) \Lambda(p,p-q,q)
 \Lambda(k-p,p-q,k-q)
 \end{eqnarray}
 
 Once again, the dressed vertex $ \Lambda $ that we
 are interested in corresponds to $ k \rightarrow 0
 $. This vertex
 scales as $ q $ on the left hand side. Power count
 of the right hand side shows that it is a linear
 function of momentum as well and thus the two sides
 are matched in exponents. To impose the amplitude
 inconsistency, we evaluate the integral on the right
 hand side in the dominant region which corresponds
 to small values of $ p $. This leads to
 
 \bea
 1 = \frac{\lambda \Lambda_0 D_1}{2 \pi
 {\Gamma_1}^2}\:\frac{2\pi}{4 \sqrt{3}}
 \eea
 
 Comparing with eqn.(13), we find
 
 \bea
 \alpha = \frac{1}{2} + \frac{1}{2 \sqrt{3} \pi}
 \simeq 0.59\
 \eea
 
 \par
 This is to be compared with the numerical value of $
 \alpha = 0.68 $. For a more careful analysis,
 eqns.(17) and (19) have to be solved numerically.
 This is an extremely formidable task because the
 dependence of $ \Lambda $ on the three variables
 (two independent) has to be charted out.
\par
As a final point, one would like to show that in this particular case, the weak scaling situation does not arise. If $ z_{\psi} $ were to be different from 2, then for $ \Sigma(k,\omega) $ to be at all relevant, $ z_{\psi} $ has to be smaller than 2.This means eqn.(12), would at zero frequency become (we now include the vertex correction)

\bea
\Sigma(k) = \frac{\lambda^2 D_1}{\Gamma_1}\:\int\:\frac{dp}{2\pi}\:\frac{\Lambda(k,p) k(k-p)}{\Sigma(k-p)}
\eea

Simple power counting shows that with $ \Lambda \propto p $, $ z_{\psi}=2 $, which contradicts our starting assumption that $ z_{\psi}<2 $ and hence there is no self-consistent solution of the weak scaling variety.
\par
We have checked to ensure that for the extended scaling case, the self-consistent scheme does give the roughening exponent. Whether, the scheme can be made to work for the weak scaling situation is under consideration.
 
 \section{Acknowledgment}
 The authors AKC and AB sincerely acknowledge partial
 financial support fom C. S. I. R., India.
 
 \section{Figure Captions}
 Fig.1a The self-consistent equation for the
 correlator with $ \it bare $ vertex. The double
 thick line is the dressed correlator and the double
 straight line the propagator. The cross stands for
 the noise.\\
 Fig.1b The self-consistent equation for the
 correlator with $ \it dressed $ vertex. The double
 thick line is the dressed correlator and the double
 straight line the dressed propagator. The cross
 stands for the noise.\\
 Fig.2 The self-consistent equation for the vertex.  
 
\end{document}